\documentclass[final, a4paper, journal, twoside]{IEEEtran}
\usepackage{fancybox}
\usepackage{xurl}
\usepackage{hyperref}
\hypersetup{colorlinks,allcolors=blue}
\usepackage{enumerate, enumitem} 
\usepackage{environ}
\usepackage{cite}
\usepackage{amsmath,amssymb,amsfonts, multirow, eurosym}
\usepackage{algorithmic}
\usepackage{graphicx}
\usepackage{textcomp}
\usepackage{textcomp, booktabs}
\usepackage[english]{babel} 
\usepackage[utf8]{inputenc} 
\usepackage{multicol} 
\usepackage{flafter} 
\usepackage{soul} 
\usepackage{color} 
\usepackage{array} 
\usepackage{tikz} 

\usepackage{glossaries}

\newcommand{\vect}[1]{\boldsymbol{#1}}
\newcommand{\ones}{\vect{1}}

\newcommand{\stateSingle}{s}
\newcommand{\state}{\vect{\stateSingle}}

\newacronym{qos}{QoS}{quality of service}
\newacronym{rdf}{RDF}{resilience density function}
\newacronym{rmf}{RMF}{resilience mass function}
\newacronym{crf}{CRF}{cumulative resilience function}

\begin{document}

\title{Resilient-By-Design: A Resilience Framework for Future Wireless Networks}

\author{Nurul Huda Mahmood, 
Sumudu Samarakoon,~\IEEEmembership{Member,~IEEE,} 
Pawani Porambage,~\IEEEmembership{Senior Member,~IEEE,} 
Mehdi Bennis,~\IEEEmembership{Fellow,~IEEE} and 
Matti Latva-aho,~\IEEEmembership{Fellow,~IEEE}
\thanks{N. H. Mahmood, S. Samarakoon, M. Bennis and M. Latva-aho are with 6G Flagship, Centre for Wireless Communications, University of Oulu, Finland. P. Porambage is with VTT, Oulu, Finland.}
\thanks{Corresponding e-mail:\href{mailto:nurulhuda.mahmood@oulu.fi}{nurulhuda.mahmood@oulu.fi}.}
}



\maketitle


\begin{abstract}
Our future society will be increasingly digitalized, hyper-connected and globally data driven. The sixth generation (6G) and beyond 6G wireless networks are expected to bridge the digital and physical worlds by providing wireless connectivity as a service to different vertical sectors, making the society increasingly dependent on wireless networks. Thus, any disruption to these networks would have a significant impact with far-reaching consequences. Disruptions can occur for a variety of reasons, including planned outages, natural disasters, and deliberate cybersecurity attacks. Resilience against such disruptions is expected to be one of the most important defining features of future wireless networks. This paper first discusses a generic framework for designing future resilient wireless networks. A novel \textbf{resilient-by-design} framework consisting of four building blocks, namely \textit{predict}, \textit{preempt}, \textit{protect} and \textit{progress}, is then proposed as a specific example. 
\end{abstract}



\section{Motivation and background}
\label{sec:intro}

\IEEEPARstart{R}{ecently} a joint statement was issued endorsing \textit{secure, open} and \textit{resilient by design} as the common design principles for the research and development of the sixth generation (6G) and beyond 6G wireless communication systems\footnote{\href{https://tinyurl.com/mr2sswnb}{Joint Statement Endorsing Principles for 6G: Secure, Open, and Resilient by Design}}. {Secure} 6G technologies are required to enable networks to fail safely and recover quickly in case of potential cybersecurity threats, while an open 6G architecture enables developing interoperable global 6G standards, interfaces, \& specifications through universally agreed processes. On the other hand, {resilience}, which is the focus of this article, is related to a system's ability to withstand a disturbance and rebound into its normal state afterwards. The ITU has formally defined it as \textit{the ability of a network or a system to continue operating correctly during and after a natural or man-made disturbance}~\cite{ITU_IMT2030}.

The term {resilience} originates from the Latin word \textit{resiliere} -- meaning \textit{to rebound or spring back} -- and is used across different academic disciplines. In the engineering domain, it is generally referred to as {a system's intrinsic ability to adjust its functionality in the presence of disturbances and unpredicted changes}~\cite{Hosseini16_resiliencyReview}. Any given network or system is subject to planned or unplanned disruptions. Most unplanned failures are induced by internal and/or external factors such as network glitches, disasters, and cyber-security attacks~\cite{RH20_resilientGuide}. Such disruptions may incur a hefty cost in terms of network downtime, lost revenues and privacy \& security breaches, especially since they are often broader in scope and more dynamic in nature than random failures.

The network's ability to avoid or cope with failures have traditionally been measured through reliability, availability and survivability~\cite{RH20_resilientGuide}. However, these metrics cannot adequately reflect the network's robustness to failures and its ability to restore normal operation following a failure. \textit{Resilience} -- which includes two aspects: maintaining quality of service (QoS) in the face of disruptions; and the quality of remediation or recovery from a failure -- is proposed as an overarching and essential design and operational characteristic of future networks that can address this gap~\cite{RH20_resilientGuide}. Resilience will be especially important in future wireless networks given that many services will increasingly be offered by mission-critical applications that will need guaranteed performance with respect to various QoS and quality of security (QoSec) metrics~\cite{mahmood21_mtcKeyDrivers}.  

\textit{Resilience} cuts through several thematic areas, such as information and network security, fault tolerance, software and hardware dependability, and network survivability. {Challenge tolerance} and {trustworthiness} are two important attributes that serve as the basis of, and is subsumed by, the concept of resilience~\cite{RH20_resilientGuide}. A challenge tolerant network is able to continue providing service in the face of internal and/or external disruptions, and includes characteristics like fault \& disruption tolerance, and survivability. On the other hand, trustworthiness -- which encompasses dependability, privacy \& security, and explainability -- guarantees that the network continues to perform as expected. 

Resilience is a rather new topic that has only been studied sporadically so far. The seminal monograph~\cite{RH20_resilientGuide} provides a comprehensive introduction to this discipline and serves as a guide to disaster-resilient communication networks in general. The concepts of resilience and mixed-criticality, which can ensure criticality awareness for diverse functional safety applications and provide fault tolerance in an autonomous manner, are jointly considered in the formulation of a resource optimization problem in~\cite{reifert2022comeback}. Therein the authors propose and evaluate four different resilience mechanisms and demonstrate considerable advantages of incorporating a mixed criticality-aware resilience strategy. Reference~\cite{khaloopour2024RBD} argues the need for future 6G networks to be resilient, and introduces a comprehensive concept for designing resilient 6G communication networks, where they suggest embedding physical and cyber resilience across all layers. On a different note, reference~\cite{Weedage2024_roamingResilience} investigates the role of national roaming as a resilience improvement approach by combining models with public data from different sources to assess the coverage and capacity of cellular networks on a country-wide scale.

Despite these (and other) notable contributions, there is still a wide gap in the existing literature on resilience from an academic perspective, especially in the context of wireless networks. This article aims to contribute towards bridging this gap by providing a comprehensive vision for resilience in future wireless networks. Our main contributions include reviewing the state of the art definition of resilience and discussing the different kinds of resiliency challenges faced by a typical wireless network. We then explore the question -- \textit{what does it mean to make a wireless network resilient?}. Finally, we present the novel \textit{resilient-by-design} concept as an specific example of a resilience framework for future wireless networks.



\section{Resilience: State of the Art Definitions and Disruption Landscape}
\label{sec:definition}

\subsection{How is resiliency conventionally defined?}
\label{sub:conventionalDef}
The general concept of resilience is defined by the system characteristics at a disruption followed by a recovery. Therein, detection of a disruption, remediation against the disturbance, and the process of recovery are pivotal for the analysis of system resilience \cite{sterbenz2010resilience}. Here, disruption can be defined as drastic changes in the system state. Suppose $\state(t)$ is the desired service level at time $t$, which is the ratio between the service state and the desired service level. For example, in a communication system, the service state could be the instantaneous data rate for a given link while the minimum rate target could be the desired level. It is worth noting that $\state(t) \succeq \ones$  implies the system meets the desired service level while the opposite indicates a detection of a disruption.

\begin{figure*}[h]
	\centering
	\includegraphics[width=0.85\textwidth]{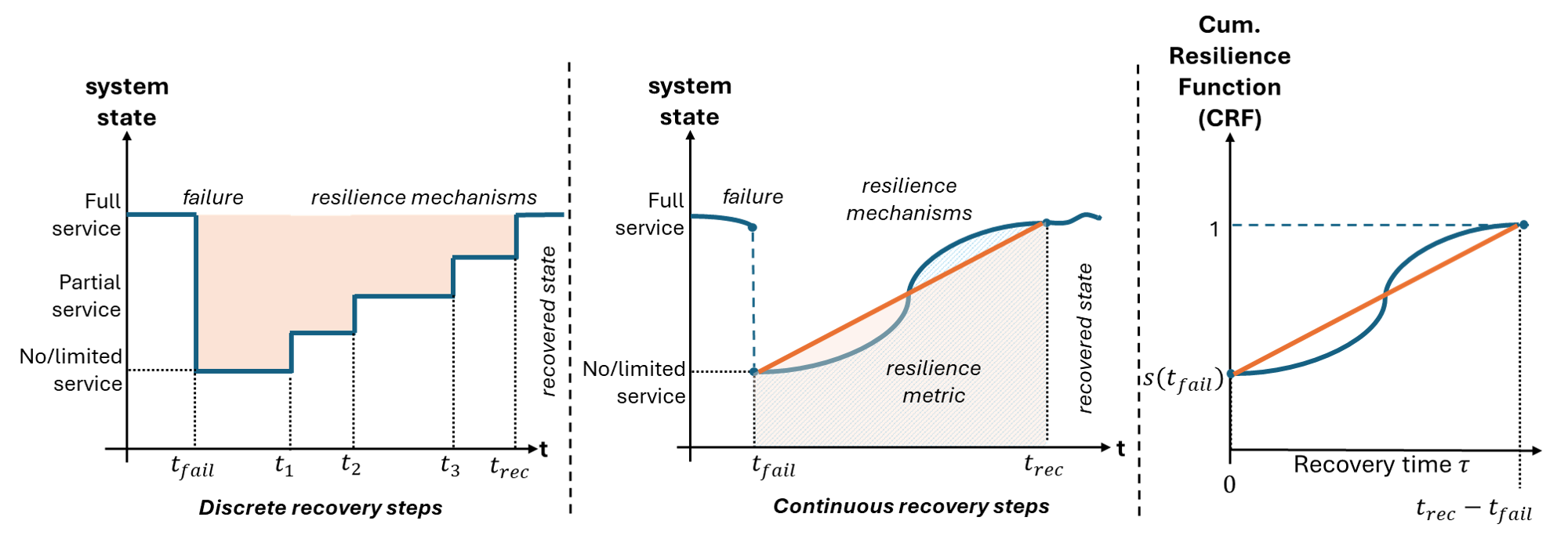}
	\caption{Illustrative example of a system's recovery process after failure through remediation with multiple recovery steps for discrete recovery steps (left), and multiple recovery mechanisms for continuous recovery steps (middle) along with their corresponding CRFs (right).}
	\label{fig:multi_adoption_recovery}
\end{figure*}

There could be several steps towards recovery and return to normal operating state once a disruption occurs. It may be in discrete steps or as different continuous processes, as illustrated in Fig. \ref{fig:multi_adoption_recovery} (left and middle, respectively). The authors of \cite{reifert2022comeback} have defined three phases for the discrete system state case, namely absorption, adoption, and recovery phase. Then, the overall system resilience is defined as the weighted sum of functions of the time spent in each phase, where the weights for each phase add up to one. Resilience in the case of continuous system state is quantified with the recovery curve \cite{sharma2018resilience}. It can be quantified using the dynamics of the recovery process as the normalized area under the system state during the recovery process, as shown by the shaded regions in Fig.~\ref{fig:multi_adoption_recovery} (middle). 

However, this definition does not distinguish between different recovery processes. For example, the two different continuous recovery processes in Fig.~\ref{fig:multi_adoption_recovery} (middle) both have the same resilience metric, though the recovery processes are different. As a remedy, the cumulative resilience function (CRF) is defined in~\cite{sharma2018resilience} as a term analogous to the cumulative density function in probability theory. The CRF, as illustrated in Fig.~\ref{fig:multi_adoption_recovery} (right), describes the system state during the recovery process, while its time derivative yields the instantaneous rate of the recovery progress which allows calculating the recovery process over any given time interval during the recovery process. Hence, these two functions allow a more detailed analysis, thereby enabling comparison between the resilience of two different recovery processes. 

\begin{figure}[htbp]
    \centering
    \includegraphics[width=0.6\columnwidth]{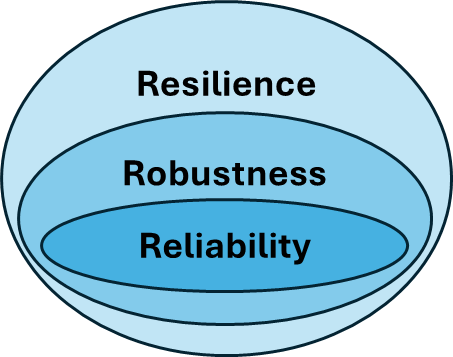}
    \caption{The relationship between reliability, robustness and resilience.}
    \label{fig:RRR}
\end{figure}

Alongside resilience, it is also worth defining the related concepts of reliability and robustness, since they are sometimes used analogously by mistake. \textit{Reliability} is the ability to perform as required for a given time interval, under given conditions, while \emph{robustness} is defined as the ability to continue operation in the face of challenges. The relation among these concepts is illustrated in Fig.~\ref{fig:RRR}. 

\subsection{Resiliency Challenges}
\label{sub:disruptions}

Wireless networks are continuously faced with a broad set of challenges, which can lead to series of cascading events resulting in link or network element failures. A challenge can result in a fault that, if not addressed, leads to a failure. For example, interference is a challenge in wireless networks. However, not all interferences result in a packet drop (fault) since wireless transmissions are made inherently robust against a certain degree of interference, e.g., through coding. Similarly, not all faults lead to failures, as some missed packets can be recovered through mechanisms such as hybrid automatic repeat request (HARQ) protocols. 

\begin{figure}[htbp]
    \centering
    \includegraphics[width=0.99\columnwidth]{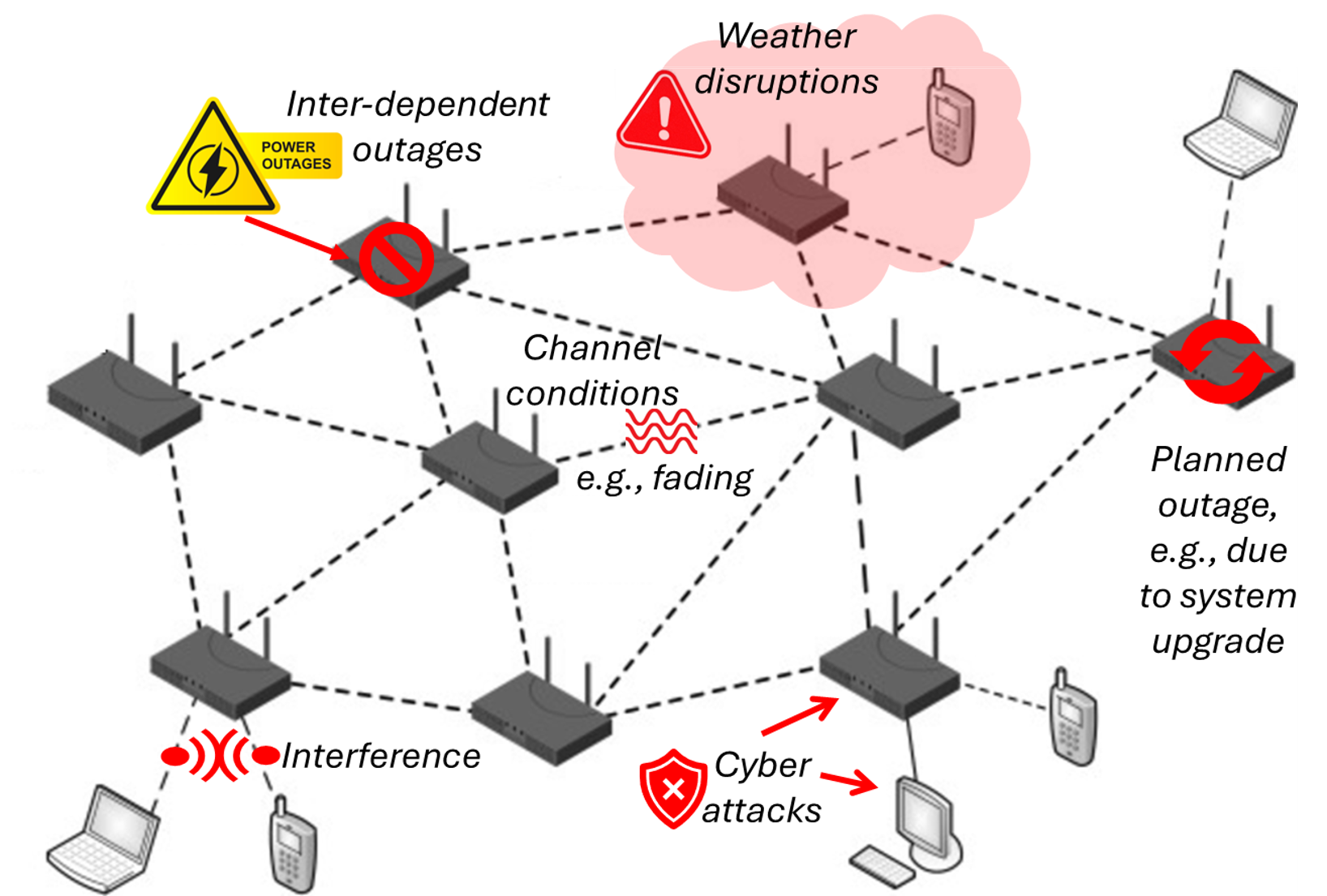}
    \caption{Different challenges that can potentially to lead to network failures affecting the resilience of a wireless network.}
    \label{fig:disruptions}
\end{figure}

Understanding the challenges to network operations that jeopardize network resilience is the first step towards designing a resilient wireless network. This section briefly discusses the different resilience challenges, as summarized in Fig.~\ref{fig:disruptions}.

\subsubsection{Internal network challenges}
\label{sub:internalDisruption}

The stochastic and shared nature of wireless transmission introduces fading and interference, which are usually the most common causes of physical layer failures. Increasing network densification further exacerbates these challenges. 
Additionally, two recent wireless networking trends may (inadvertently) expose the network to new sources of disruptions. The first is the move towards openness. While this promotes interoperability, it also opens a new level of complexity and introduces compatibility challenges that may expose the network to multiple points of failures. A related but different challenge arises from the pervasive use of generative artificial intelligence (AI) in network design and software development. Ensuring the quality and reliability of such generated code is a primary challenge as they can introduce new sources of network failures such as logical errors that are much more difficult to identify and debug.

\subsubsection{Unintentional external factors}
\label{sub:unintentionalExternal}

A significant part of unplanned failures are induced by unintentional external factors such as natural disasters and adverse weather effects~\cite{RH20_resilientGuide}. These generally affect the lower layers (network layer and below). Such failures are often much broader in scope and impact than internal failures. Unfortunately, disaster induced network failures will be more prevalent in the future as climate change increases the frequency and intensity of extreme weather events. Disruptions may also occur due to technical failures in interdependent networks such as power or backhaul networks and due to solar storms~\cite{machua2016_techDisasters}. Although not a network failure per se, the lack of network availability in remote areas also affects resilience as it impacts the key 6G value of \textit{inclusiveness} and challenges the global coverage target and the use of public wireless networks to serve critical communications\footnote{For example, \href{https://www.ericsson.com/en/cases/2025/ericsson-and-erillisverkot-the-next-chapter}{the TETRA based public safety communications network in Finland will be replaced by a system operating on Elisa's (a Finnish operator) commercial 4G/5G network}}.

\subsubsection{Resilience against cyber-security threats}
\label{sub:externalThreat}

Digital infrastructure, including wireless networks, are targeted by cyber-attacks more than ever before not only for financial gains, but also with malicious intentions. Such attacks are diverse in terms of the layer it targets, ranging from the physical layer (e.g., jamming attacks) to the application layer (e.g., exploit attacks). The instability of the geo-political situation and technological advancements have amplified the frequency, intensity, and impact of such attacks. AI advancements have become a double-edged sword in this regard. They enable autonomous security management while also contributing to the rise of sophisticated AI-driven attacks~\cite{BT20_AIfor5Gsecurity} even though regulatory frameworks such as GDPR and the European AI Act pose strict requirements on data protection and ethical use of AI in general. It is also important to make networks resilient against physical layer security attacks such as electromagnetic pulse attack, attacks against the control channel, eavesdropping, authentication attacks, and reconfigurable intelligent surface (RIS)-induced attacks. Finally, quantum technology driven security attacks are foreseen as another pressing threat. The capability of quantum computing to break classical cipher keys and the conventional encryption methods used in current systems will lead to unprecedented vulnerabilities in wireless networks. 

\section{Resilience in Wireless Networks}
\label{sec:resiliencyInWireless}

The ultra reliable low latency communications (URLLC) service class introduced in fifth generation (5G) New Radio paved the door towards providing connectivity as a service for different vertical sectors. 
Although URLLC presented a paradigm shift in wireless network evolution, it nevertheless failed to holistically provide the desired connectivity solution for mission critical services due to several limitations~\cite{park2020extreme}. Firstly, URLLC is application agnostic whereas different mission critical applications have different design requirements. Secondly, while reliability and latency are important key performance indicators (KPI), there are also other important metrics like the distribution of failures or jitter that need to be considered. Next, URLLC is inherently limited in scale as it only focuses on a single link. Finally, security and trustworthiness considerations are not reflected. To address these shortcomings, resilience has recently been proposed as a key characteristic of future wireless networks~\cite[\& references therein]{reifert2022comeback, khaloopour2024RBD}, including by the International Telecommunications Union (ITU)~\cite{ITU_IMT2030}.

Within different engineering disciplines, attempts have been made to mathematically define resilience, e.g.,~\cite[\& references therein]{RH20_resilientGuide, Hosseini16_resiliencyReview}. However, these definitions should be revisited and revised to reflect the specific characteristics of wireless networks and the mission-critical applications it aims to support in the future. Furthermore, given the importance of safeguarding against security and trust vulnerabilities discussed in Section~\ref{sub:externalThreat}, such considerations must be a prominent feature of resiliency in future wireless networks. 

\begin{figure}[thb]
    \centering
    \includegraphics[width=0.99\columnwidth]{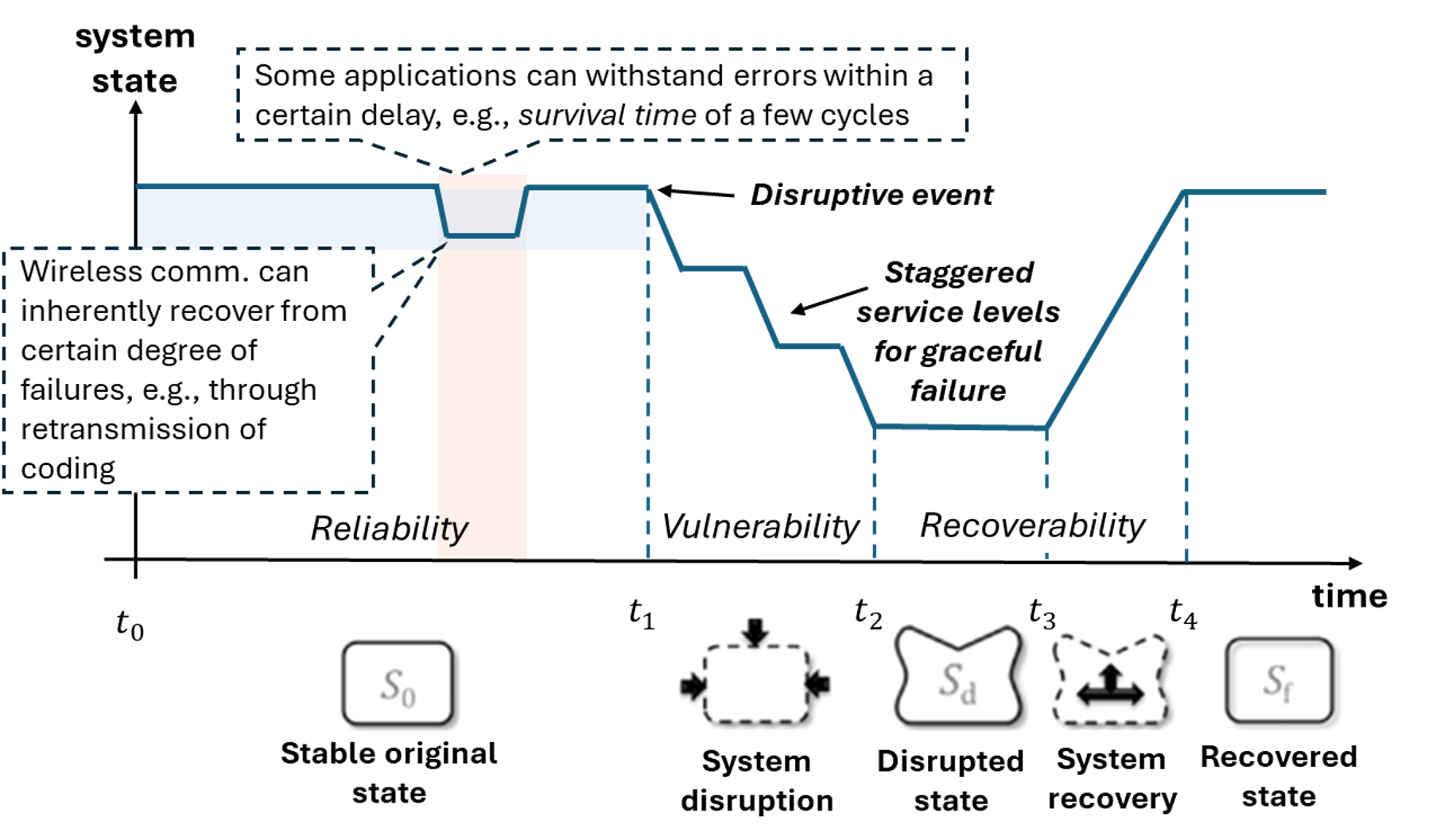}
    \caption{Resiliency definition with respect to the particular characteristics of wireless communication.}
    \label{fig:resiliencyDef}
\end{figure}

\begin{figure}[t]
    \centering
    \includegraphics[width=0.85\columnwidth]{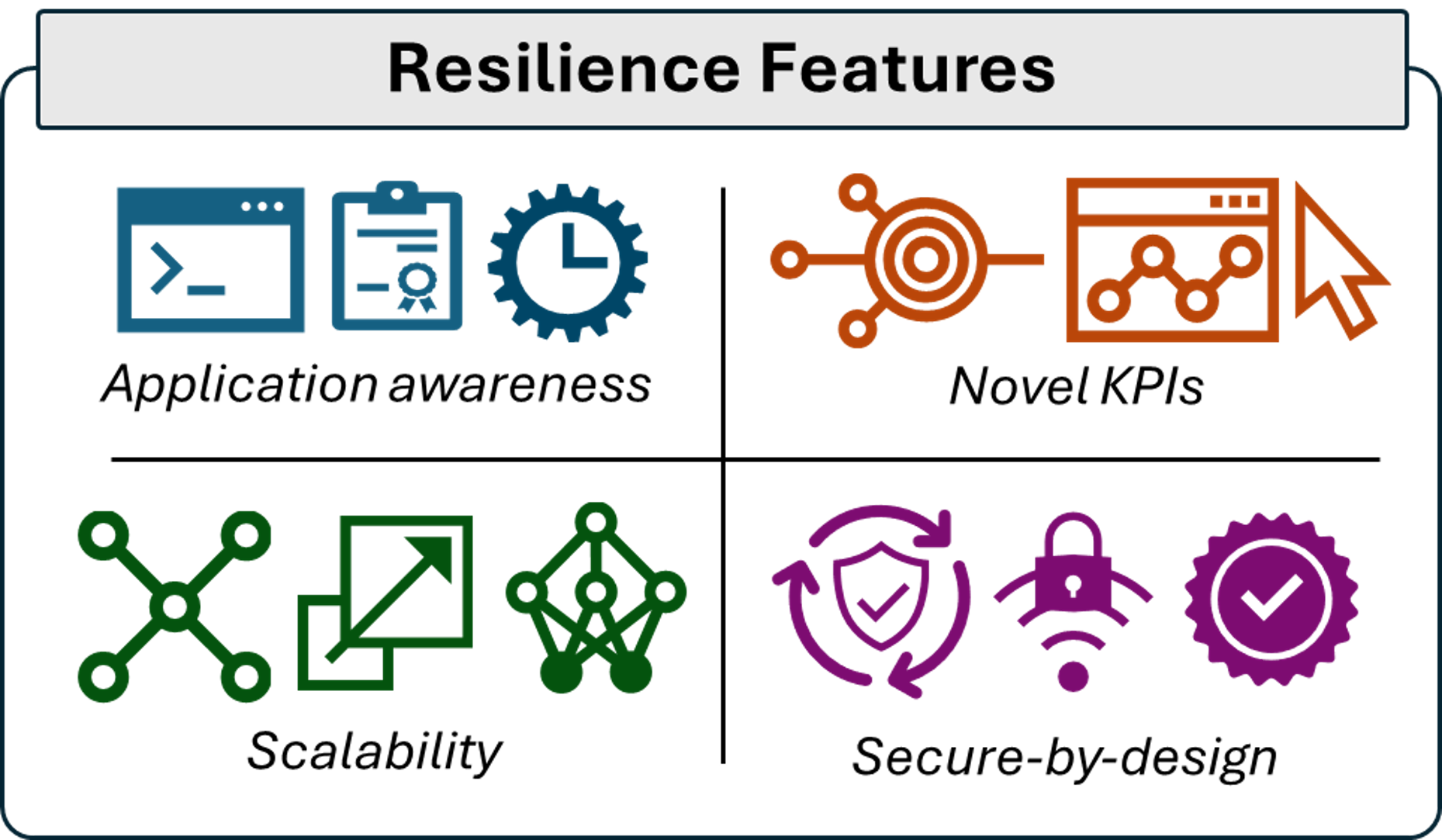}
    \caption{Key characteristics of a resilient wireless network.}
    \label{fig:rbdConcept}
\end{figure}

The very nature of wireless propagation makes it vulnerable to random fading, interference and other fluctuations. Hence, wireless systems have always been designed to inherently tolerate certain degrees of failures, as exemplified in Fig.~\ref{fig:resiliencyDef}.  A simple example is the HARQ protocol which allows wireless transmitters to retransmit a packet in the event of a transmission failure. However, a resilient wireless system should not only be able to address disruptions due to the stochastic nature of the wireless propagation channel, but also withstand faults and errors arising from other sources described in Section~\ref{sub:disruptions}. In the rest of this section, we suggest four key characteristics of a resilient wireless networks, as illustrated in Fig.~\ref{fig:rbdConcept}.

\subsection{Application-awareness}
Different applications have different requirements, some of which reflect the particular nature of wireless communications. For instance, machines in an industrial process have a survival time (time without receiving new data) of several cycle times (update interval of data) before which it would stop and move to a safe position. 5G systems introduced a new QoS architectural model to enable QoS-aware resource allocation. More specifically, two QoS flows are defined: Guaranteed Bit Rate (GBR) and non-GBR flows. GBR flows guarantees a minimum bit rate along with a hard upper bound, whereas the default non-GBR flow do not provide any rate-related guarantees. However, this only provides guarantees on the bit rate, does not work well under unpredictable traffic conditions, and is agnostic to new requirements for 6G. Hence, a resilient network should be able to differentiate different applications according to their requirements and serve them accordingly. This can be done, for example, through the adoption and widespread communication of service level agreements (SLA) along with service requests~\cite{alevizaki2023_distributedService}. The SLAs would define the priority, quality, and responsibility associated with the requested service. 

Under normal operating condition, the network efficiency can be optimized by allocating resources to applications based on the SLAs instead of supporting a generic reliability or latency target, as in the case of current URLLC service class. SLAs can also ensure the continuity of service by defining staggered service levels, and allowing prioritizing certain applications over others during disruption situations where network resources are constrained. As an example, an operator may commit to provide a certain level of service under normal operating conditions, while providing reduced service levels under different disruption situations.  

\subsection{Key Performance and Value Indicators (KPIs and KVIs)}

Efficiency will be an important cornerstone of future resilient networks. Efficient optimization of different resilience aspects require defining novel KPIs and metrics that can facilitate mathematical formulation of the resiliency optimization problem. Reliability and latency are the main metrics measuring a wireless network's/link's ability to cope with failures in URLLC. Resilience has a wider scope and hence must considers other metrics as well. For example, in the case of process automation described earlier with a given survival time, the distribution of failures, the number of consecutive failures, and the mean time between failures are more relevant metrics than reliability~\cite{MAM+22_missionEffCap}. It will also be important to consider the time needed to return to normal service after disruption. Collating this with differentiated service levels under disruptions, the rebound rate need to be separately measured for each level (please see Figs.~\ref{fig:multi_adoption_recovery} and~\ref{fig:resiliencyDef}). Alongside, future wireless networks including 6G will not only achieve the performance requirements set by the applications/use cases but also create values to the society in different forms measured through the novel concept of key value indicators (KVI). 

\subsection{Scalability}
Differently from URLLC, which focuses on a single point-to-point transmission link, resilience is a system wide concept that must consider a multi-hop network and define methods to enable and evaluate resiliency over a multi-hop communication link. Hence, resilience can be considered as the ability of a network as a whole, as opposed to a single link, to mitigate and swiftly recover from the disruptions caused by various reasons. As an illustrative example, suppose we have a wireless network of $N$ nodes. A resilient network should aim to maximize the number of nodes that can fail before the overall network connectivity breaks, minimize the vulnerability of key node and reduce the dependency among the nodes without compromising the overall network efficiency in terms of resource utilization and redundancy provisioning. This will become increasingly important as network densifies, i.e. $N$ increases.

\subsection{Secure-by-Design}

The \textit{secure-by-design} concept is also a core resilience requirement which aims to proactively address cyber-security threats and privacy breaches. The fundamental security features such as the CIA triad (i.e. confidentiality, integrity, availability), and AAA framework (i.e., authentication, authorization, accounting) need to be well preserved in the design phase of the hardware and software components, the communication links, the protocols, and the interfaces in cyber-physical systems. However, the evolving threat surface in the communication networks and the enabling technologies are creating unprecedented challenges in ensuring secure-by-design concepts in next- generation networks. In the design of network topology, data management frameworks, or application scenarios, it is critical to integrate continuous risk assessment, rigorous security testing, sustainability, and adaptability to ensure cyber-resilience and enhance cyber-threat intelligence.

\section{Resilient-by-Design: A Proposed Resilience Framework}
\label{sec:resiliencyDesign}

\begin{figure*}[htbp]
    \centering
    \includegraphics[width=0.7\textwidth]{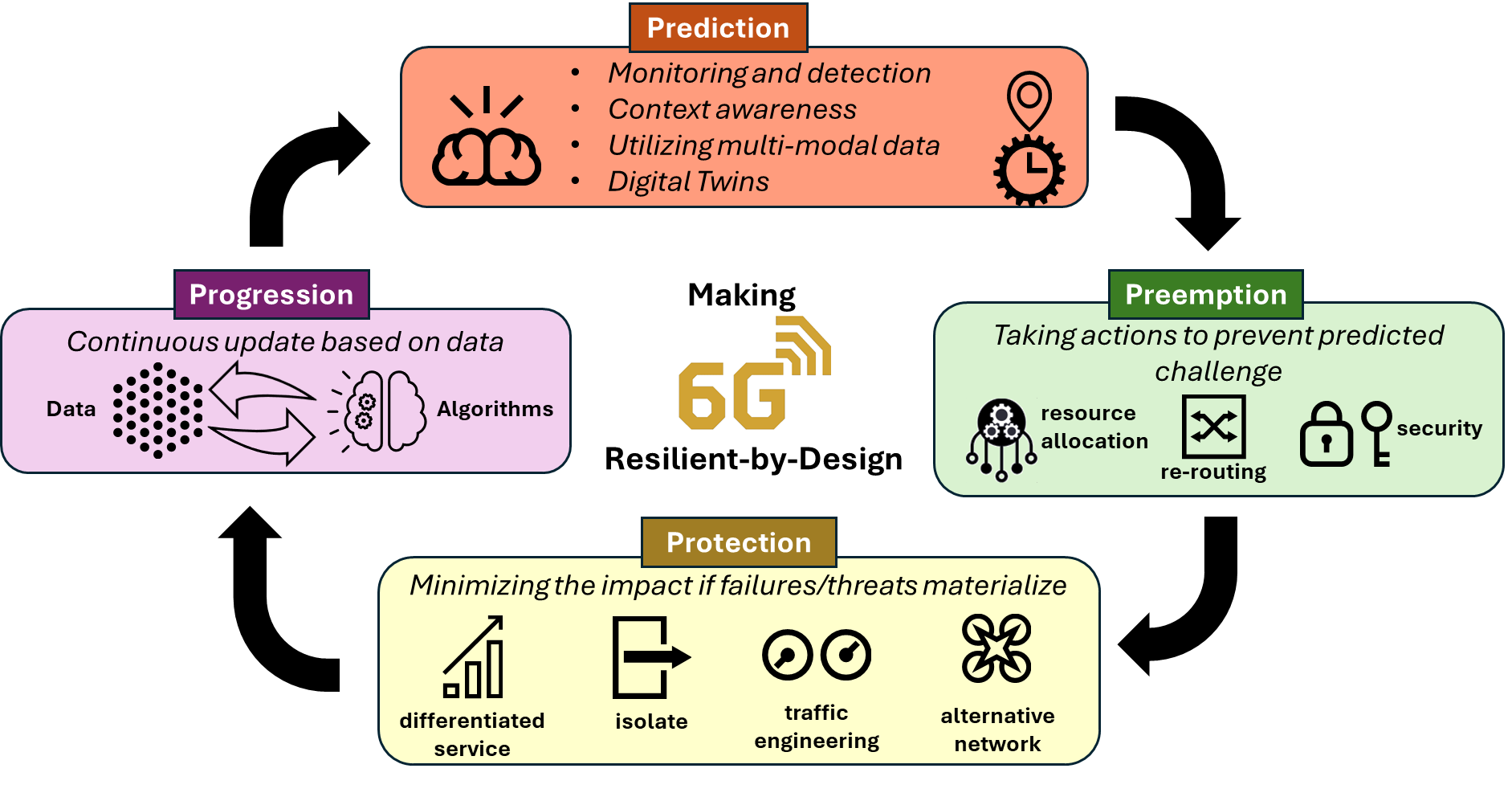}
    \caption{Illustration of the proposed resilient-by-design concept for 6G networks.}
    \label{fig:resilient6G}
\end{figure*}

This section proposes a novel \textit{resilient-by-design} concept consisting of four building blocks, as illustrated in Fig.~\ref{fig:resilient6G}, as a specific example of the resilience framework introduced in Section~\ref{sec:resiliencyInWireless}. 
It may be highlighted that resilience is a broad concept covering a wide spectrum that cannot be covered within the limited scope of this article. Hence, the enablers outlined here are by no means exhaustive in depth and breadth.

\subsection{Predicting failures and threats}
\label{sub:predict}

Wireless networks are already becoming capable of autonomously self-configuring, self-optimizing and self-healing based on real-time conditions. Ensuring resilience requires going beyond that to be able to continuously monitor, detect and predict failures, external disruptions and security threats using model-based or data-based approaches. 

For the former, detailed modeling of the fading and/or interference correlation can be utilized to build a Bayesian framework to predict future channel/interference conditions. In addition, context-awareness by utilizing external or auxiliary information and fusing multi-modal data can further enhance the prediction accuracy. Examples include the use of weather forecast to predict adverse weather events, using information of planned large events to predict network traffic surge, or even the use of social media data to identify events in real time that will have a significant impact on the network operation. Context awareness can also be utilized to assess security threats, anticipate potential security risks, and identify the required security level~\cite{chorti2022_contextAwareSecurity}.

Data-driven AI methods offer a wide array of tools to anticipate issues such as link failures, node outages, congestion, and cyber-attacks like jamming or spoofing. Supervised learning models can classify known failure types or regress performance degradation, while unsupervised methods, such as clustering or autoencoders, are effective in detecting anomalies in unlabeled data. Time-series models like long short-term memory (LSTMs) and temporal convolutional networks forecast performance and behavioral trends, enabling proactive interventions in failures and disruptions. Graph-based approaches, particularly Graph Neural Networks, model the network topology to predict cascading failures and structural vulnerabilities. Reinforcement learning, both model-free and model-based, supports adaptive decision-making under uncertainty. From a security point of view, it is important to consider training models that are not prone to adversarial attacks when using AI tools.

Additionally, the integration of digital twins with AI enables real-time simulation and stress-testing of failure scenarios, offering predictive insights, enabling more accurate tracking and prediction of the propagation environment and thus, facilitating system reconfiguration. Adversarial training is a specific example of the application of digital twins in the security domain. Moreover, RISs have opened the paradigm of engineering the propagation environment, thereby transforming the random wireless environment into a controllable channel that makes it easier to monitor and detect failures.

\subsection{Preemption}
\label{sub:preempt}

Once a challenge is predicted, preemptive actions are required to address them. Examples of such preemptive actions at the link level include proactive resource allocation, centralized or distributed coordination schemes and redundancy measures. In the case of redundancy schemes, the inefficiency associated with it can be addressed by provisioning redundancy such that it is activated only when predicted to be needed. Different traffic flows with different SLAs can also be intelligently multiplexed to further enhance the efficiency. Predicted node failures on a larger scale can be addressed by proactive rerouting of the traffic. In this case, the correlation / dependency structure between the different possible routes, and causal relation with contextual information need to be considered to minimize the probability of the predicted failure impacting the rerouted path(s). For example, if a failure is expected due to weather disruption in a given area, the rerouted path should avoid that area. 

On the security front, preemptive actions are needed to mitigate potential security breaches. Predictive analytics allows recognizing anomalous behavior, which can proactively identify and counter potential threats. Alongside, AI/ML tools can help in mitigating potential security risks by identifying network or data anomalies and defining optimum countermeasures. In addition to identifying known attacks as mentioned above, more sophisticated mechanisms are also required to predict, detect and mitigate unknown attacks or zero-day attacks which are becoming more common. As the conventional approach, 5G considers perimeter-based security where the network is enclosed within a trusted zone to prevent unauthorized access. However, with open interfaces, disaggregation of hardware and software components, and the dynamic nature of complex 6G and beyond network architecture, it is impossible to clearly demarcate a network perimeter to define static trust zones and security controls as countermeasures. Zero Trust Architecture and adaptive security models which require authentication, authorization and continuous validation for all devices can be applied under such circumstances. 

\subsection{Protection}
\label{sub:protect}

In the unwanted case that a failure or security breach occurs, immediate actions must be taken to address the disruption and minimize its impact on the network. The following is a list of potential solutions that can be adopted as countermeasures.

\textbf{\textit{Isolation}}: future 6G networks will be a network of networks, with many special purpose sub-networks interconnected with each other. The overall network architecture must be agile enough to isolate and remove the affected subnetwork(s) to minimize the disruptions and safeguard the rest of the network from the detrimental impact of the failure. At the same time, critical network functionalities should be distributed and across the network with sufficient redundancy to minimize the impact of removing one of more subnetworks on the operation of the remaining network. 

\textbf{\textit{Staggered service level}}: the network should define different operating modes with different SLAs defined for different operating conditions or system states. This will ensure service continuity even when the system does not meet the desired service level (ie., $\state(t) \prec \ones$, albeit offering a different (often lower) service level. In this case, critical and delay-sensitive services will need to be prioritized. 

\textbf{\textit{Traffic engineering}}: networks usually witness a surge in the network traffic in the event of a disruption as users try to access information, which further strains the network. Traffic engineering can be applied to proactively divert the traffic to the unaffected parts of the network or even block traffic from less critical applications based on the SLAs. Furthermore, proactively broadcasting information updates to potential information-seeking users can prevent possible traffic surges. 

\textbf{\textit{Novel network architecture}}: major disruptive events like earthquakes can affect the network architecture, making part of the network unavailable. Such challenges can be addressed by utilizing novel network architectures such a space, air and ground integrated networks (SAGIN) which integrates satellites (also known as non-terrestrial network (NTN)), high altitude platform stations and unmanned aerial vehicles with the terrestrial network (TN). The integration of SAGIN with TNs can provide continuity in service or minimize the disruption in case of such failures, though the heterogeneity of the different networks is a major challenge as these networks may be designed with different and not complementary principles and approaches regarding security and resilience. It may be noted that TN-NTN integration is part of the 3GPP standardization roadmap for 5G and 6G, which partially addresses some of these interoperability issues.

\subsection{Progression}
\label{sub:progress}

The final step of the resilient-by-design concept does not work in real-time but rather is a process that continuously works in the background. This step's goal is to continuously upgrade and update the resilience mechanisms in the other three steps. For the \textit{prediction} part, this involves updating the underlying model parameters or even adopting new models or tools based on new training data to increase the prediction accuracy. For example, incorporating continual learning or transfer learning can help towards achieving this in the context of AI/ML tools. Similarly, the probabilities of variables in Bayesian Networks establishing the causal relationship in the network should be updated as new observations are added to the model. 

Continuous monitoring of KPIs of interest in the \textit{pre-emption} and \textit{protection} steps can help in evaluating the effectiveness of the adopted preemptive strategies and comparing among different options. Alongside, the cost in terms of resource and/or energy usage must also be monitored. This will allow selecting strategies that meet the desired performance criteria while providing the optimum cost-performance tradeoff. For instance, joint security-QoS optimization and continuous adaptation of the security protocols can be considered for fine-tuning security mechanisms to achieve the best performance without compromising protection. This includes balancing the load on network elements for security operations, ensuring efficient resource usage, and minimizing latency introduced by the security processes.


\section{Conclusions and Way Forward}
\label{sec:conclusions}

Resilience is expected to be one of the most important and defining features of 6G and beyond 6G networks since future networks must be made resilient to provide and maintain an acceptable level of service in the face of disruptions and challenges to normal operation. The sources of disruption include challenges related to fading and interference stemming from the stochastic and shared nature of wireless propagation, disruptive events like extreme weather conditions and natural disasters, challenges in providing ubiquitous connectivity in remote and sparsely populated areas or when the existing infrastructure is unavailable, and attacks by malicious actors. 

This article first introduces the important and emerging concept of resilience by outlining several key characteristics of a resilient wireless network that can address the aforementioned challenges. Thereafter, we have proposed a novel four-stage resilience framework called the \textit{resilient-by-design} concept. We believe that a network can be made resilient if it is able to \textit{predict} disruptive events, make \textit{preemptive} decisions to mitigate the disruption, and take action to \textit{protect} and minimize its impact on the network. Alongside, the network must \textit{progress} to update and upgrade its defense mechanisms by continuously learning from the current operating conditions. Further research challenges related to resilience lie ahead with important open questions to be addressed, for example, how to measure resilience, what metrics should be used, and how to assign numerical values for those? 

\section*{Acknowledgments}
{We thank the anonymous reviewers whose comments and suggestions helped improve and clarify this manuscript. This work was supported by the European Union’s Horizon Europe research and innovation programme within \href{https://hexa-x-ii.eu/}{Hexa-X-II project} (grant no. 101095759), and the Research Council of Finland through the projects 6G Flagship (grant no. 369116) and ReWIN-6G (grant no. 357120).}

\bibliographystyle{IEEEtran}



\vspace{-10mm}
\begin{IEEEbiographynophoto}
{Nurul Huda Mahmood} is a senior researcher and Adjunct Professor at Center for Wireless Communications (CWC), University of Oulu (UOulu), and the coordinator of Wireless Connectivity research area in \href{https://www.6gflagship.com/}{6G Flagship}. His current research focus is on resilient wireless networks.
\end{IEEEbiographynophoto}
\vspace{-10mm}
\begin{IEEEbiographynophoto}
{Sumudu Samarakoon [M]} received his B. Sc. degree in Electronic and Telecommunication Engineering from the University of Moratuwa, Sri Lanka in 2009, the M. Eng. degree from the Asian Institute of Technology, Thailand in 2011, and Ph. D. degree in Communication Engineering from UOulu in 2017. He is currently working in CWC as an assistant professor.
\end{IEEEbiographynophoto}
\vspace{-10mm}
\begin{IEEEbiographynophoto}
{Pawani Porambage [SM]} Pawani Porambage is a senior scientist at VTT technical research centre of Finland. She is also an adjunct professor at the UOulu. Her research interests include security aspects of wireless networks.
\end{IEEEbiographynophoto}
\vspace{-12mm}
\begin{IEEEbiographynophoto}
{Mehdi Bennis [F]} is a Professor at CWC, UOulu, where he heads the intelligent connectivity and networks/systems group (ICON). His main research interests are in radio resource management, heterogeneous networks, game theory, and distributed ML in wireless networks. He has been the recipient of several prestigious awards including the 2015 Fred W. Ellersick Prize from the IEEE Communications Society, the 2016 Best Tutorial Prize from the IEEE Communications Society, and the 2020 Clarviate Highly Cited Researcher by the Web of Science. 
\end{IEEEbiographynophoto}
\vspace{-12mm}
\begin{IEEEbiographynophoto}
{Matti Latva-aho [F]} received the M.Sc., Lic.Tech. and Dr. Tech (Hons.) degrees in Electrical Engineering from the UOulu in 1992, 1996 and 1998, respectively. Currently he is the Vice-Rector for research at the University of Oulu, and a professor on wireless communications at CWC. He is also a Global Research Fellow with Tokyo University. His research interests are related to mobile broadband communication in 6G systems. He is a recipient of the Nokia Foundation Award in 2015.
\end{IEEEbiographynophoto}

\end{document}